\documentclass[preprint,onecolumn,aps,nofootinbib]{revtex4}
\usepackage{graphicx}%Include figure files
\pdfoutput=1
\usepackage[colorlinks=true,linkcolor=blue,urlcolor=blue,filecolor=black,citecolor=red,pdfstartview=FitV,
pdftitle={},pdfsubject={},pdfkeywords={},pdfpagemode=None,bookmarksopen=true]{hyperref}
\usepackage{epsfig}
\usepackage{amsmath}
\usepackage{amsfonts}
\usepackage{amssymb}
\usepackage{color}%
\usepackage{dcolumn}% Align table columns on decimal point
\providecommand{\U}[1]{\protect\rule{.1in}{.1in}}

\newcommand{\f}{\begin{equation}}
\newcommand{\ff}{\end{equation}}
\newcommand{\fa}{\begin{eqnarray}}
\newcommand{\ffa}{\end{eqnarray}}

\allowdisplaybreaks[4]
\begin{document}
\title{Note on the butterfly effect in holographic superconductor models}
\author{Yi Ling $^{1,3,4}$}
\email{lingy@ihep.ac.cn}
\author{Peng Liu $^{1}$}
\email{liup51@ihep.ac.cn}
\author{Jian-Pin Wu $^{2,3}$}
\email{jianpinwu@mail.bnu.edu.cn}
\affiliation{$^1$ Institute of High Energy Physics, Chinese Academy of Sciences, Beijing 100049, China\ \\
$^2$ Institute of Gravitation and Cosmology, Department of
Physics,
School of Mathematics and Physics, Bohai University, Jinzhou 121013, China\ \\
$^3$ Shanghai Key Laboratory of High Temperature Superconductors,
Shanghai, 200444, China\ \\
$^4$ School of Physics, University of Chinese Academy of Sciences,
Beijing 100049, China}
\begin{abstract}
In this note we remark that the butterfly effect can be used to
diagnose the phase transition of superconductivity in a
holographic framework. Specifically, we compute the butterfly
velocity in a charged black hole background as well as
anisotropic backgrounds with Q-lattice structure. In both
cases we find its derivative to the temperature is discontinuous
at critical points. We also propose that the butterfly
velocity can signalize the occurrence of thermal phase transition
in general holographic models.
\end{abstract}
\maketitle

\section{Introduction}
Recently quantum butterfly effect has been becoming a hot
spot of research which links the gauge/gravity duality to quantum
many-body theory and quantum information theory
\cite{Maldacena:1997,Witten:1998,
Roberts:2016wdl,Shenker:2013pqa,Shenker:2013yza,Roberts:2014isa,Roberts:2014ifa,Shenker:2014cwa,Kitaev:2014v1,
Maldacena:2015waa,Polchinski:2015cea,Hosur:2015ylk,Polchinski:2016xgd,
Swingle:2016var,Blake:2016wvh,Blake:2016sud,Lucas:2016yfl,Ling:2016ibq,Fan:2016ean,Alishahiha:2016cjk,Yao:2016ayk,Garttner:2016mqj,zhai:2016exp,Feng:2017wvc}.
Diagnosed by the out-of-time-order correlation (OTOC) functions,
the butterfly effect describes the information scrambling over
a quantum chaotic system. On gravity side, the butterfly
effect is described by a shock wave geometry on
the horizon that can be induced by an infalling particle
which is exponentially accelerated. The butterfly effect
ubiquitously exists in holographic theories due to its sole
dependence on the near horizon data of the gravitational bulk
theory. In particular, the Lyapunov exponent $\lambda_L$ is always
characterized by the Hawking temperature of the black hole as
$\lambda_L=2\pi k_BT$, while the butterfly velocity is completely
determined by the horizon geometry
\cite{Roberts:2016wdl,Blake:2016sud,Ling:2016ibq}. Moreover, a
bound on chaos is proposed as $\lambda_L \leqslant 2\pi k_BT$ and
the saturation of this bound is viewed as the criterion for a
quantum chaotic system to have a classical gravity dual
description \cite{Maldacena:2015waa}. Stimulated by above investigation in holographic
approach, many physicists in condensed matter as well as quantum
information community have made great efforts in the
measurement of the OTOC in laboratory
\cite{Swingle:2016var,Yao:2016ayk,Garttner:2016mqj,zhai:2016exp}.
The related progress is supposed to provide more practical
tools to test the proposals in holographic theories, and in turn
push forward the investigation on butterfly effects in quantum many-body systems.

In recent paper \cite{Ling:2016ibq} we have investigated the
butterfly effect in holographic models which exhibit
metal-insulator transition (MIT) and found that the butterfly velocity
$v_B$ can diagnose quantum phase transitions (QPT). The key
point on this is that the occurrence of QPT usually
involves the RG flows from UV to different IR fixed points \cite{Donos:2012js}.
On the other hand, the butterfly velocity $v_B$ depends on
the IR geometry solely. Therefore, the change of IR fixed points
may be reflected by the distinct behavior of $v_B$. In this
note we intend to argue that the butterfly effect can exhibit
attractive behavior during the course of thermal phase transition
as well. This extension is natural, since based on
Landau theory the occurrence of thermal phase transition is always
accompanied by a symmetry breaking characterized by some order
parameter. While in the context of holography, the spontaneous
breaking of symmetry is usually a reflection of the instability of
the background in bulk, signalized by the appearance of black hole hair
which is supposed to deform the horizon, namely IR geometry
strongly. Therefore, to provide evidence to support above
argument we will investigate the temperature behavior of the
butterfly velocity in holographic superconductor models.
Specifically we will demonstrate that the derivative of $v_B$ to
the temperature is discontinuous at critical points of phase
transition.

We organize this paper as follows. In next section we will
first consider the butterfly effect in the simplest holographic model
with superconductivity which is constructed over a charged black
hole. Then we turn to study this effect over more complicated
backgrounds with lattice structure in subsection \ref{scqlat}. A brief
discussion about possible extensions and experimental prospects will be
presented in the end of this note.

\section{Butterfly effects and holographic superconductivity}
In this section we will first introduce the butterfly velocity on anisotropic background.
Using the anisotropic butterfly velocity results, we reveal that the butterfly velocity could diagnose
superconductivity phase transitions.
\subsection{Butterfly velocity on anisotropic background}\label{vb-aniso}
Given a background with a black brane, we can compute the shockwave solution
on the horizon generated by a particle freed at the asymptotic AdS region. The butterfly
effect is represented by this sort of shockwave geometry, from which the Lyapunov exponent
and butterfly velocity can be read off \cite{Ling:2016ibq,Blake:2016sud,Blake:2016wvh,Roberts:2016wdl}.
For a generic anisotropic black brane geometry,
\fa && ds^2={1\over z^2}\Big[-(1-z)f(z)dt^2+\frac{dz^2}{(1-z)f(z)}+V_x(z)dx^2+V_y(z)dy^2\Big]
\label{bs}
\ffa
the butterfly velocity is also anisotropic, which can be written as \cite{Ling:2016ibq}
\fa
\bar{v}_B(\theta)  =\left. v_B \sqrt{\frac{\sec ^2(\theta )
V_x\left(z\right)}{V_x\left(z\right)+\tan ^2(\theta )
V_y\left(z\right)}}\right|_{z=1}\,,
\label{vbtheta}
\ffa
where $\theta$ is the polar angle. $v_B = \bar v_B (0)$ is the
butterfly velocity along the $x$-direction, given by
\fa
v_B=
\left.\sqrt{\frac{-2\pi \hat T
V_y(z)}{V_y(z)[V'_x(z)-2V_x(z)]+V_x(z)[V'_y(z)-2V_y(z)]}}
\right|_{z=1}\,, \label{vbx}
\ffa
where the prime denotes the derivative to the radial coordinate $z$, and the $\hat T$ is the Hawking temperature of black brane (\ref{bs}).
The metric ansatz (\ref{bs}), (\ref{vbtheta}) and (\ref{vbx}) are applicable for subsequent two holographic models in next two subsections.

Next, we investigate the butterfly effects in two holographic models involving phase transitions of superconductivity.
\subsection{Butterfly effects in a simple holographic superconductor}\label{vb-rn}
The minimal ingredients to build a superconductor model in a holographic
framework are provided by adding a charged complex scalar field into
Einstein-Maxwell theory, in which the Lagrangian is
\cite{Hartnoll:2008vx,Hartnoll:2008kx}
\fa \label{action-SI}
\mathcal{L}_I=R+6 -\frac{1}{4}F^2 -|D_{\mu}\Psi|^2-M^2|\Psi|^2\,.
\ffa

Notice that we have set the AdS radius $L=1$. $F=dA$ is the
curvature of $U(1)$ gauge field $A$. $\Psi$ is the charged
complex scalar field with mass $M$ and scaling dimension
$\Delta_{\Psi}=3/2+(9/4+M^2)^{1/2}$, and the charge $q$.
$D_{\mu}=\partial_{\mu}-iqA_{\mu}$ is the covariant derivative.
We solve (\ref{action-SI}) by taking metric ansatz (\ref{bs}) and
\fa
A=\mu(1-z)a(z) dt\,, \qquad \qquad \Psi=\Psi(z),
\label{bs1}
\ffa
where $f(z)\equiv(1+z+z^2-\mu^2z^3/4)S(z)$ and $\mu$ is the
chemical potential in dual field theory. The Hawking temperature
is then given by
\fa \hat{T}=\frac{(12-\mu^2)S(1)}{16\pi}\,.
\label{temperature}
\ffa
The corresponding dimensionless temperature is $T=\hat{T}/\mu$.
Note that the system (\ref{action-SI}) is isotropic, the anisotropic
metric (\ref{bs}) could be limited as isotropic case, \emph{i.e.}, $V_x = V_y$.
For simplicity,
we set the mass and charge of the complex scalar field as $M^2=-2$
and $q=2$ such that its scaling dimension is $\Delta_{\Psi}=2$
and its asymptotical behavior at UV is
\fa
\Psi=z\Psi_1+z^2\Psi_2\,. \label{psiUV}
\ffa
We shall treat $\Psi_1$ as the source and $\Psi_2$ as the expectation value
of the dual operator. At the same time, we set $\Psi_1=0$
such that the condensation will take place spontaneously. At
high temperature, the solution to equations of motion with
ansatz (\ref{bs}) is simply the Reissner-Nordstr\"om AdS (RN-AdS)
black brane solution with $\Psi(z)=0$ and $S(z)=a(z)=V_x(z)=V_y(z)=1$. However, below the
critical temperature, imposing regular boundary conditions on the
horizon and requiring the scalar field to decay at UV as in
Eq.(\ref{psiUV}), one can numerically find new black
brane solutions with scalar hair, which is dual to a
superconducting phase in the boundary theory.

\begin{figure}
\center{
\includegraphics[scale=0.7]{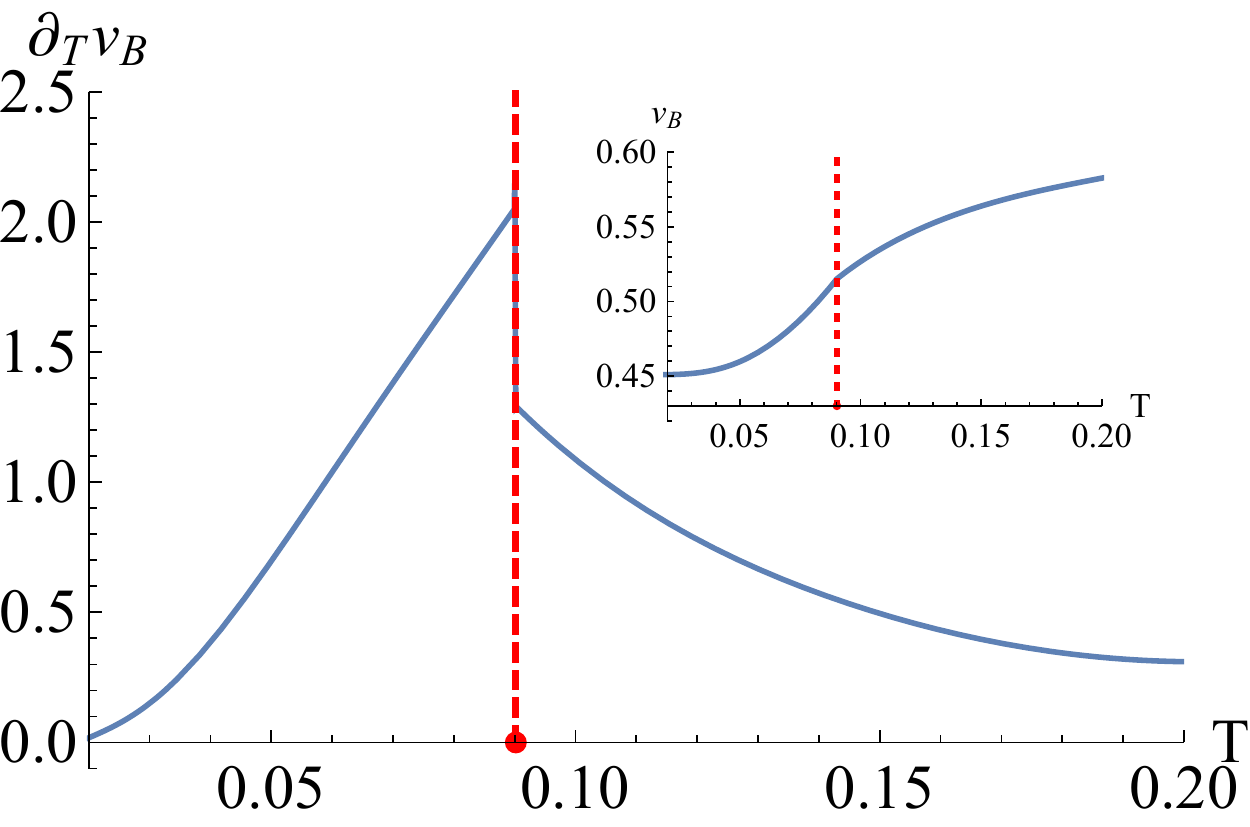}\ \\
\caption{\label{vbrn1} Plot of $\partial_Tv_B$ as a function of temperature $T$ in the holographic superconductor model (\ref{action-SI}).
The inset plot shows $v_B$ as a function of $T$.
The vertical dashed line denotes the superconducting phase transition temperature $T_c$, which is $T_c\simeq0.0902$.
}}
\end{figure}

In this simple model with isotropy we have $V_x(z)=V_y(z)$ such that the
formula is simplified as $v_B=\sqrt{\pi T\mu/[2V_x(1)-V'_x(1)]}$. Now we
numerically compute $v_B$ as the function of temperature $T$
during the course of phase transition. Fig.\ref{vbrn1} shows
$\partial_Tv_B$ as a function of temperature $T$ and the inset
plot shows $v_B$ vs. $T$. In this figure it is evident that
the derivative $\partial_Tv_B$ is discontinuous at the
critical temperature $T_c$ (the red dashed line in vertical direction),
which indicates that the butterfly velocity can be utilized
as a new independent probe of the phase structure of the superconductor.

%%%%%%%%%%%%%%%%%%%%%%%%%%%%%%
%%%%%%%%%%%%%%%%%%%%%%

\subsection{Butterfly effects in holographic Q-lattice superconductor}\label{scqlat}
The second model we consider is the holographic
superconductor on Q-lattices, which has been studied in
\cite{Ling:2014laa}. Its Lagrangian reads as
\fa
\mathcal{L}_Q=R+6-\frac{1}{4}F^2 -|D_{\mu}\Psi|^2-M^2|\Psi|^2 -|\nabla\Phi|^2-m^2|\Phi|^2.
\label{action-SQ-v}
\ffa
In comparison with the
Lagrangian in (\ref{action-SI}), an additional neutral complex
scalar field $\Phi$ with mass $m$ is introduced to break the
translational invariance \cite{Donos:2013eha}. We can solve this
gravitational system by taking the metric (\ref{bs}) and
\fa
\Phi=e^{i \hat{k}x}z^{3-\Delta_{\Phi}}\phi(z)\,,\qquad A=\mu(1-z)a(z) dt\,, \qquad \Psi=\Psi(z),
\label{Phi-ansatz}
\ffa
where $\Delta_{\Phi}=3/2+(9/4+m^2)^{1/2}$
is the scaling dimension of $\Phi$.
Note that since we only introduce the lattice in $x$-direction, our geometry is anisotropy.
Now, each black brane
solution is characterized by three scaling-invariant parameters,
\emph{i.e}., the Hawking temperature $T\equiv\hat{T}/\mu$, the
lattice amplitude $\lambda\equiv\hat{\lambda}/\mu^{3-\Delta_{\Phi}}$ with
$\hat{\lambda}=\phi(0)$, and the wave vector $k\equiv\hat{k}/\mu$.
For normal states ($\Psi=0$), there exist MITs when adjusting $\lambda$ or $k$ \cite{Donos:2013eha}.
A complete phase diagram can be found in
\cite{Ling:2014laa,Ling:2015dma}. While the superconducting phase
has been numerically found in \cite{Ling:2014laa}. It has been
shown in \cite{Ling:2014laa} that the lattice structure
suppresses the condensation and the critical temperature is lowered compared
with the situation when the lattice is absent. The phase structure in \cite{Ling:2014laa}
demonstrates that the transition to the superconducting
phase from the metallic phase is easier than from the insulating phase.

\begin{figure}
\center{
\includegraphics[scale=0.55]{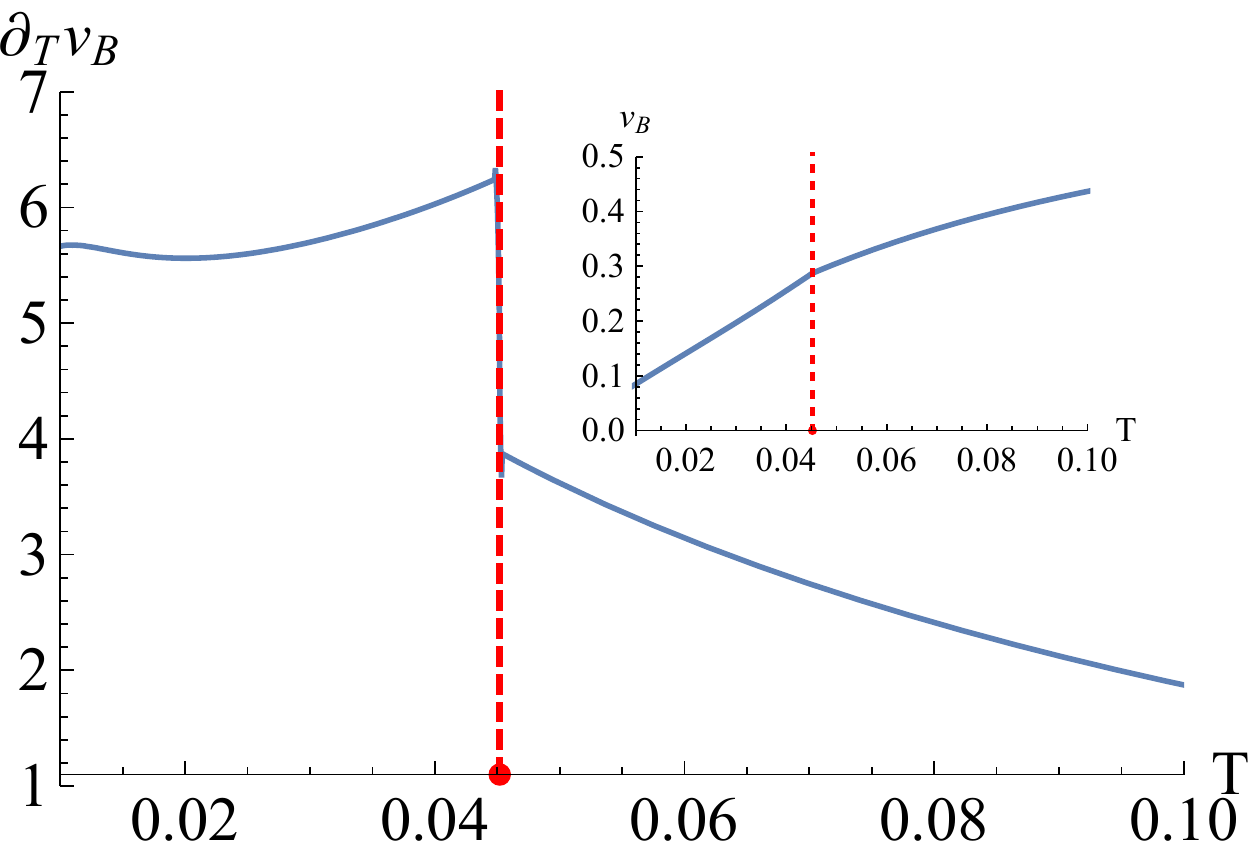}\ \hspace{0.8cm}
\includegraphics[scale=0.55]{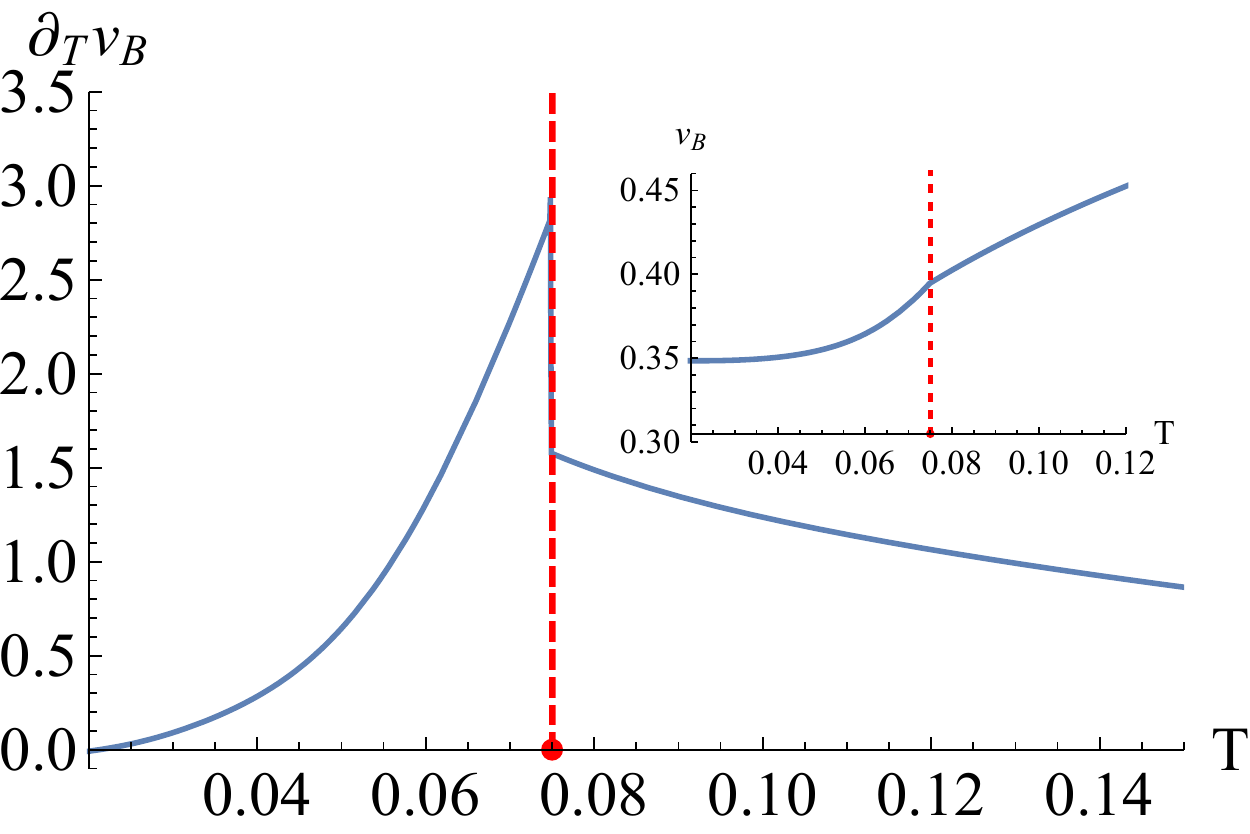}\ \\
\caption{\label{vbQ} Plots of $\partial_Tv_B$ as a function of temperature $T$ in holographic Q-lattice superconductor.
The inset plots show $v_B$ as a function of $T$.
The vertical dashed line denotes the superconducting phase transition temperature $T_c$.
Left plot is for $\lambda=2$ and $k=0.5$, which is insulating phase and the critical temperature $T_c\simeq 0.0452$.
While right plot is for $\lambda=2$ and $k=1.5$, which is metallic phase and the critical temperature $T_c\simeq 0.0750$.
}}
\end{figure}

\begin{figure}
\center{
\includegraphics[scale=0.65]{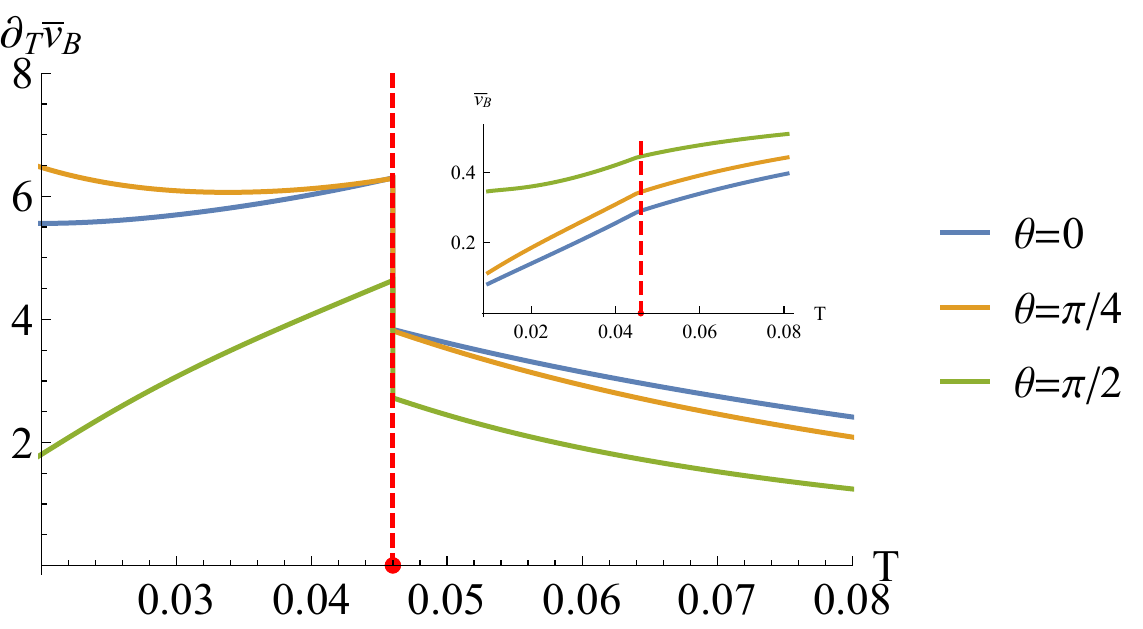}\ \hspace{0.8cm}
\includegraphics[scale=0.65]{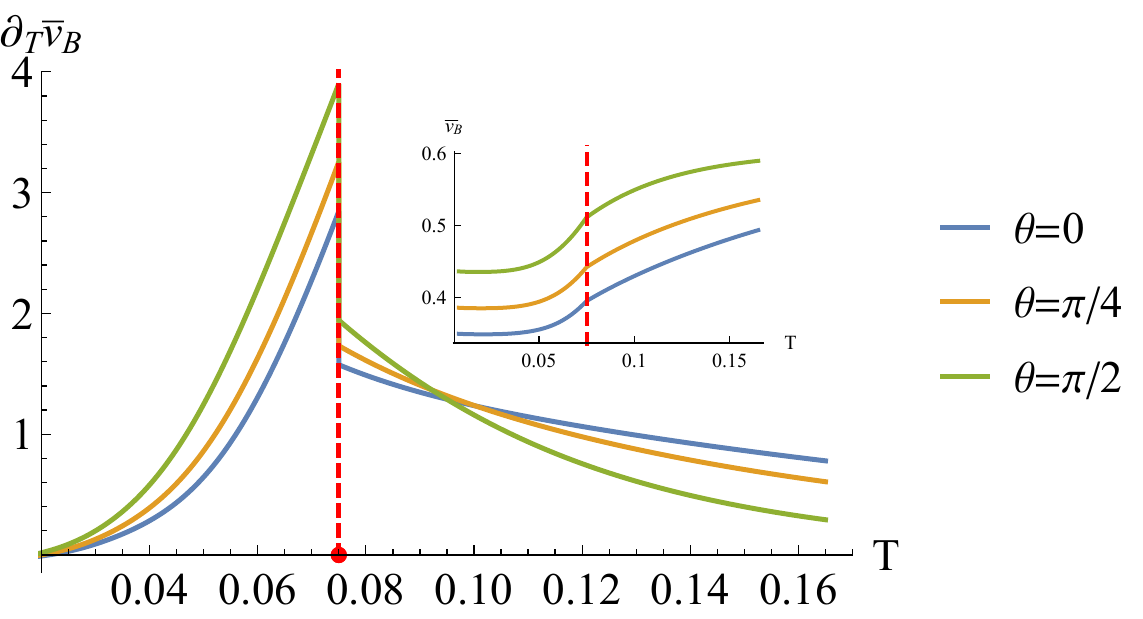}\ \\
\caption{\label{vbQa} Plots of $\partial_T\bar{v}_B$ as a function of temperature $T$ in holographic Q-lattice superconductor
at different directions.
The inset plots show $\bar{v}_B$ as a function of $T$.
The vertical dashed line denotes the superconducting phase transition temperature $T_c$. The parameters in
both plots have the same values as those in Figure (\ref{vbQ}) correspondingly.
}}
\end{figure}

\begin{figure}
\center{
\includegraphics[scale=0.55]{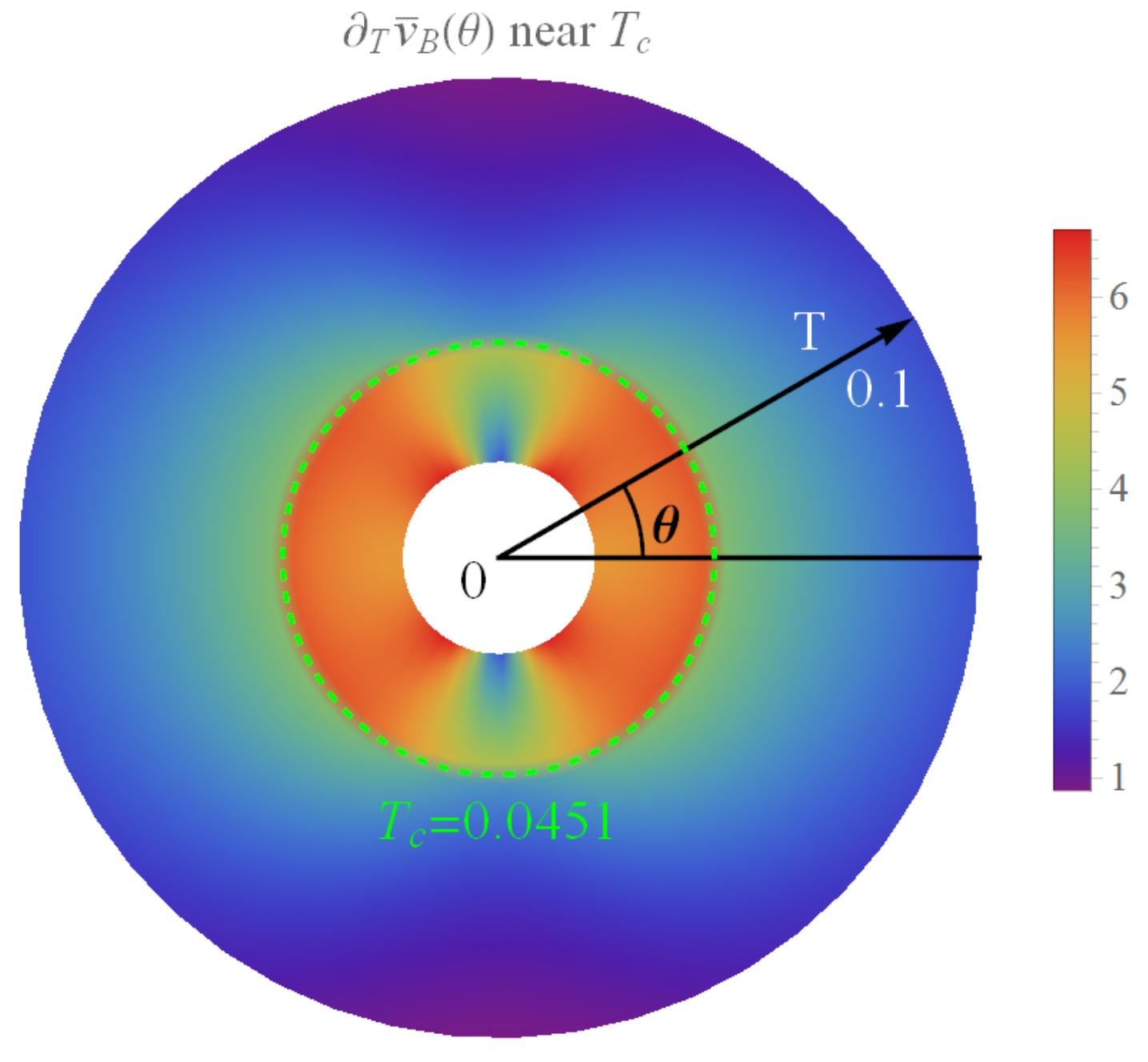}
\caption{\label{disc}
The density plot of angle dependence of $\partial_T \bar v_B(\theta)$ near the critical point $T = T_c$ at parameter $\lambda=2, \, k=0.5$.
The radial direction is $T$, plotting between $0.02<T<0.1$ with $T_c \simeq 0.0451$ (the green dotted circle).
}}
\end{figure}

Now our main purpose is to compute the butterfly velocity $\bar v_B(\theta)$ for a
given background and then observe its behavior during the phase
transition from a normal phase which could be metallic or
insulating to a superconducting phase.
First, we focus on the butterfly velocity along the $x$-direction,
\emph{i.e.},  $\bar v_B(0)=v_B$. For simplicity, here we set $\{M^2,m^2,q\}=\{-2,-2,2\}$ and
demonstrate the temperature behavior of $v_B$ in two typical cases
in Fig.\ref{vbQ}, one for metal-superconductor transition and the
other for insulator-superconductor transition. In both cases we
find that the first order derivative of $v_B$ with respect to temperature is
discontinuous. Next we examine the butterfly
velocity $\bar{v}_B$ as the function of temperature $T$ in
different directions (see Fig.\ref{vbQa}). It is interesting to notice that at a given temperature,
the butterfly velocity increases with the increasing polar angel in the first quadrant
\footnote{The period of $\bar{v}_B(\theta)$ is $\pi$.},
which implies the lattice structure suppresses the propagation of quantum information.
We demonstrate the anisotropy of $\partial_T \bar v_B(\theta)$ more transparently in Fig.\ref{disc}, which
is on account of the introduction of the lattice structure. It can be seen from Fig.\ref{disc} that
$\partial_T \bar v_B(\theta)$ is discontinuous at $T_c \simeq 0.0451$ in any direction. This reflects
the fact that the emergence of the condensation, which is responsible
for the discontinuity of $\partial_T \bar v_B(\theta)$, depends solely on the temperature. Also,
the period of $\partial_T \bar v_B(\theta)$ can be clearly read off as $\pi$, respecting the period of $\bar v_B(\theta)$.
In summary, the butterfly velocity is a good diagnose of the superconducting phase transition.

%%%%%%%%%%%%%%%%%%%%%%%%%%%%%%
%%%%%%%%%%%%%%%%%%%%%%
%%%%%%%%%%%%%%%%%%%%%%%%%%%%%%
%%%%%%%%%%%%%%%%%%%%%%

%%%%%%%%%%%%%%%%%%%%%%%%%%%%%%
%%%%%%%%%%%%%%%%%%%%%%
%%%%%%%%%%%%%%%%%%%%%%%%%%%%%%
\section{Discussion}
In this note we have proposed that the butterfly effect
should exhibit distinct behavior during the course of thermal
phase transition. In our two holographic superconductor models
we have explicitly demonstrated that the first
derivative of $v_B$ to temperature is discontinuous.

Next we point out some possible generalizations of our work.
First, it is desirable to analytically obtain the discontinuity of
$\partial_T \bar v_B(\theta)$ in two models we discussed. Second,
based on the arguments presented in the introduction,
it is natural to expect that the interesting phenomenon in this note
can be observed in other holographic superconductor models.
In addition, we also expect that the butterfly effect can capture the
occurrence of other sorts of thermal phase transition as well, for
instance, the transition between RN black holes and the dilatonic
black holes \cite{Cadoni:2009xm}, or the MIT induced by Charge Density Waves (CDW)
\cite{Ling:2014saa}. Finally, we conjecture that the non-analytical
behavior of the butterfly velocity at critical temperature
could be diverse. For example, instead of the first-order discontinuity that
we have observed in this note, zero-order discontinuity or higher-order discontinuity
might be observed in other sorts of thermal phase transitions.

Furthermore, we expect what we have observed for the
butterfly effect in holographic context can be extended to
characterize thermal phase transition in a realistic system which
may not have a gravity dual. The validity of such an extension in
principle could be tested experimentally in light of recent
progress on the measurements of the OTOC \cite{Swingle:2016var,Yao:2016ayk,Garttner:2016mqj,zhai:2016exp}.

\begin{acknowledgments}

We are very grateful to Meng-he Wu, Zhuo-yu Xian,
Yi-kang Xiao and Xiang-rong Zheng for helpful discussion. This work is supported by the
Natural Science Foundation of China under Grant Nos.11275208, 11305018 and 11575195,
and by the grant (No. 14DZ2260700) from the Opening Project of Shanghai Key
Laboratory of High Temperature Superconductors. Y.L. also acknowledges the support from Jiangxi
young scientists (JingGang Star) program and 555 talent project of
Jiangxi Province. J. P. Wu is also supported by the Program for
Liaoning Excellent Talents in University (No. LJQ2014123).
\end{acknowledgments}

\end{document}